\documentclass[showpacs,preprintnumbers,aps]{revtex4}
\begin{document}
\preprint{USM-TH-154}
\title{Fractional statistics and confinement}
\author{Patricio Gaete}
\email {patricio.gaete@usm.cl} \affiliation{Departamento de
F\'{\i}sica, Universidad T\'ecnica F. Santa Mar\'{\i}a,
Valpara\'{\i}so, Chile}
\author{Clovis Wotzasek}
\email {clovis@if.ufrj.br}
\address{Instituto de F\'\i sica, Universidade Federal do Rio de
Janeiro, 21945, Rio de Janeiro, Brazil}
\date{\today}
\begin{abstract}
It is shown that a pointlike composite having charge and magnetic
moment displays a confining potential for the static interaction
while simultaneously obeying fractional statistics in a pure gauge
theory in three dimensions, without a Chern-Simons term. This
result is distinct from the Maxwell-Chern-Simons theory that shows
a screening nature for the potential.
\end{abstract}
\maketitle

The explanation for the fact that the fundamental constituents
of matter are confined is a long-standing theoretical physics problem.
Its solution has evaded full comprehension up to now and a great
many deal of progress has been made in order to distinguish
between the apparently related phenomena of screening
and confinement. In fact the distinction between screening and
confinement is of considerable importance in our present
understanding of gauge theories. In order to avoid the
complexities of four dimensions the strategy of many of these
studies has been the restriction to lower dimensions as a
theoretical laboratory.

In this work we show that a pointlike composite having charge and
magnetic-dipole is able to provide a clear-cut realization of
anyon fractional statistics in (2+1) dimensions while displaying a
confining feature among two such composites. The anyonic
fractional statistics \cite{Wilczek} displayed by this system is realized without
the introduction of a Chern-Simons term, this way avoiding the
appearance of topological mass that is responsible for the
screening characteristics displayed by pure electric charges in
the electrodynamics controlled by the Maxwell-Chern-Simons theory
\cite{Elcio}.  Differently, the electrodynamics of the composite
studied here is a pure Maxwell theory. In this case a rich
physical structure was found to exist and the issue of confinement
has been satisfactorily settled. The full understanding is obtained by
studying the potential created by two oppositly charged composite
some distance apart.

Electrodynamics in (2+1) dimensions have been extensively
discussed in the last few years\cite{Deser,Dunne,Khare}. It is
well known that in three dimensions it is allowed the possibility
of particles with any statistics, where the physical excitations
obeying it are called anyons. A concrete way to realize
non-trivial statistics is by attaching a magnetic flux to
electrically charge particles. A simple way to attach a flux to an
elementary charge is to use the Chern-Simons theory to control
the dynamics of the photons so that Wilczek's charge-flux
composite model of the anyon can be implemented\cite{Wilczek}.

However, it should be recalled that when the Chern-Simons term is
added to the usual Maxwell term, the gauge field becomes massive,
leading to topologically massive gauge theories. Consequently
electric and magnetic fields are virtually screened at distances
bigger than $\Lambda \approx 1/m$ and the statistics of anyons
changes with their mutual distance in the presence of the Maxwell
term \cite{Shisuya}. However, it has been shown possible to
describe anyons without the introduction of Chern-Simons
term\cite{Itzhaki, Gaete3} through the introduction of a
generalized connection allowing this way the realization of
fractional statistics.

In the present context it is also important to recall that the
interaction potential between static charges is an object of
considerable interest.  In fact the ideas of screening and
confinement play a central role in modern gauge theory and the
physical content can be understood when a correct separation of
the physical degrees of freedom is made \cite{Gaete} using a
formalism in terms of physical (gauge-invariant) quantities. This
method has been used previously for studying features of screening
and confinement in two-dimensional quantum electrodynamics,
generalized Maxwell-Chern-Simons gauge theory and for the
Yang-Mills field \cite{Gaete}. After the dynamical feature of the
composite is presented below, this method will be applied to study
the interaction energy of two such entities.

As already stated, our objective is to compute explicitly the
interaction energy between static pointlike sources for this new
electrodynamics. For this purpose we shall first carry out its
Hamiltonian analysis.
The gauge theory we are considering is defined by the following
Lagrangian in three-dimensional space-time:
\begin{equation}
{\cal L} =  - \frac{1}{4}F_{\mu \nu }^2  - A_\mu  J^\mu .
\label{gen1}
\end{equation}
For a static composite locate at $\bf x_0$, the ``charge-magnetic
dipole" is defined by $ J_0 = \rho$ and $J_k =
\frac{\mu}{e}\epsilon_{km}\partial_m \rho$ and has been introduced
before by Deser \cite{Deser-GA} in another context, where $\rho =
e {\delta^{(2)} \left( {\bf x - \bf x_0}\right)}$ and $\mu$ is the
dipole's moment \cite{Colatto}. For a composite located at the origin, the
solution of the Maxwell equations read,
\begin{equation}
B\left( {\bf x} \right) = \mu \delta ^{\left( 2 \right)} \left(
{\bf x} \right), \label{ecua3}
\end{equation}
\begin{equation}
E_i \left( {\bf x} \right) =  - \frac{e}{{2\pi }}\frac{{x^i
}}{{r^2 }}, \label{ecua4}
\end{equation}
with $ r \equiv |\bf x|$. One immediately sees that the associated
magnetic field has its support only at the position of the
composite, and a long range electric field is also generated. As a
consequence, the total magnetic flux associated to the magnetic
field is
\begin{equation}
\Phi _B  = \int\limits_V {d^2 } xB\left( x \right) = \mu .
\label{ecua5}
\end{equation}
This tells us that the charged dipole actually behaves like a
magnetic flux point. As it turns out the composite can realize
fractional statistics without introducing a Chern-Simons term.
This provides a novel way to describe anyons. Accordingly, the
mechanism of attaching a magnetic flux to the charges has been
implemented in a particularly simple way, as was claimed in Ref.
\cite{Itzhaki}.

Next let us consider the screening versus confinement question. It
is well known that under the Maxwell-Chern-Simons dynamics, the
static potential for two oppositely charged particles is given by
\begin{equation}
V =  - \frac{{e^2 }}{\pi }K_0 \left( \eta |\bf y - \bf y^ \prime|
\right), \label{potz}
\end{equation}
where $K_0$ is a modified Bessel function and $\eta$ is a massive
cutoff. This result displays the screening character of the MCS
potential for anyons.

To obtain the corresponding behavior for the charge-magnetic
dipole composite we must carry out the quantization of the system.
The Hamiltonian analysis of the system starts with the computation
of the canonical momenta $ \Pi ^\mu   = F^{\mu 0}$, which produces
the usual primary constraint $\Pi ^0=0$ and $\Pi ^i  = F^{i0} $
($i=1,2$ ) while the canonical Hamiltonian has the usual
maxwellian structure
\begin{equation}
H_C  = \int {d^2 x} \left\{ { - \frac{1}{2}F_{i0} F^{i0}  +
\frac{1}{4}F_{ij} F^{ij}  - A_0 \left( {\partial _i \Pi ^i  - J^0
} \right)   + A_i J^i } \right\}. \label{gen2}
\end{equation}
Preservation in time of the primary constraint leads to the
secondary constraint $\Omega _1 \left( x \right) \equiv \partial
_i \Pi ^i \left( x \right) - J^0 \left( x \right)=0$ and together
displays the first-class structure of the theory. Using $ c_0
\left( x \right)$ and  $ c_1 \left( x \right)$ as Lagrange
multiplier fields to implement the constraints, the corresponding
total (first class) Hamiltonian that generates the time evolution
of the dynamical variables then reads
\begin{eqnarray}\label{gen3}
H &=& H_C  + \int {d^2 x} \left\{ {c_0 \left( x \right)\Pi _0
\left( x \right) + c_1 \left( x \right)\Omega _1 \left( x \right)}
\right\}\nonumber\\
&=& \int {d^2 x} \left\{ { - \frac{1}{2}F_{i0} F^{i0}  +
\frac{1}{4}F_{ij} F^{ij}  + c^\prime  \left( x \right)\left(
{\partial _i \Pi ^i  - J^0 } \right)   + A_i J^i } \right\}
\end{eqnarray}
where $ c^{\prime }(x)=c_{1}(x)-A_{0}(x)$. We have used that $\Pi
^0 =0$ for all time and  $ \dot{A}_0 \left( x \right) = \left[
{A_0 \left( x \right),H} \right] = c_0 \left( x \right)$, which is
completely arbitrary, to eliminate $ A^0 $ and $ \Pi^0 $.

To fix the gauge it is convenient to choose
\begin{equation}
\Omega _2 \left( x \right) \equiv \int\limits_{C_{\xi x} } {dz^\nu
} A_\nu  \left( z \right) = \int_0^1 {d\lambda } x^i A_i \left(
{\lambda x} \right) = 0 \label {gen5}
\end{equation}
as a constraint, where  $\lambda$  $\left( {0 \le \lambda  \le 1}
\right)$ is the parameter describing the spacelike straight path
between the reference point $ \xi ^k $ and $ x^k $ , on a fixed
time slice. For simplicity we have assumed the reference point
$\xi^k=0$. The choice (\ref{gen5}) leads to the Poincar\'{e} gauge
\cite{Gaete}. Through this procedure, we arrive at the following
set of Dirac brackets for the canonical variables
\begin{equation}
\left\{ A_{i}(x),A^{j}(y)\right\} ^{*}=0,  \label{gen6}
\end{equation}
\begin{equation}
\left\{ \pi _{i}(x),\pi ^{j}(y)\right\} ^{*}=0,\label{gen6b}
\end{equation}
\begin{equation}
\left\{ A_{i}(x),\pi ^{j}(y)\right\} ^{*}=g _{i}^{j}\delta
^{(2)}\left(
x-y\right) -\partial _{i}^{x}\int_{0}^{1}d\lambda \text{ }%
x^{j}\delta ^{(2)}\left( \lambda x-y\right) ,  \label{gen7}
\end{equation}
while the Dirac brackets of the magnetic $\left( B=\varepsilon
_{ij}\partial ^{i}A^{j}\right)$ and electric $( E^i  = \Pi ^i )$
fields are the canonical ones:
\begin{equation}
\left\{ {E_i \left(  x \right),E_j \left(  y \right)} \right\}^
* = 0 , \label{gen8}
\end{equation}
\begin{equation}
\left\{ {B \left(  x \right),B \left(  y \right)} \right\}^
* = 0 , \label{gen9}
\end{equation}
\begin{equation}
\left\{ {E_i \left( x \right),B\left(  y \right)} \right\}^
* = - \varepsilon _{ij} \partial _x^j \delta ^{(2)} \left( x - y
\right). \label{gen10}
\end{equation}
It is important to realize that, unlike the Maxwell-Chern-Simons
theory, in the present model, the brackets (\ref{gen6}) and
(\ref{gen8}) are commutative. The equations of motion for the
electric and magnetic fields are,
\begin{equation}
 {\dot E}_i( x) =  - \varepsilon _{ij}
\partial ^j B\left(  x \right)  + J_i \left(  x \right) + \int {d^2 y}
J^k \left(  y \right)\partial _k \int\limits_0^1 {d\lambda } y_i
\delta ^{(2)} \left( {\lambda  x -  y} \right), \label{gen11}
\end{equation}
\begin{equation}
{\dot B}\left(  x \right) =  - \varepsilon _{ij} \partial _i E_j
\left(  x \right). \label{gen12}
\end{equation}
In the same way, we write the Gauss law as:
\begin{equation}
\partial _i E_L^i    - J^0  =
0, \label{gen13}
\end{equation}
where $E_L^i$ refers to the longitudinal part of $E^i$. These are
the Ampere, Faraday and Gauss laws, respectively. Together they
imply that for a static composite located at $x^i=0$, the static
electromagnetic fields are given by Eqs. (\ref{ecua3}) and
(\ref{ecua4}). The set of Dirac brackets above is then elevated to the category quantum commutators as usual \cite{Dirac1}.

After achieving the quantization we may now proceed to calculate
the interaction energy between pointlike sources in the model
under consideration. To do this, we will compute the expectation
value of the energy operator $H$ in a physical state $ \left|
\Omega \right\rangle$. We also recall that the physical states $
\left| \Omega  \right\rangle$ are gauge-invariant \cite{Dirac}. In
that case we consider the stringy gauge-invariant $\left|
{\overline \Psi  \left( \bf y \right)\Psi \left( {\bf y^ \prime }
\right)} \right\rangle$ state,
\begin{equation}
\left| \Omega  \right\rangle  \equiv \left|\overline \Psi \left(
{\bf y} \right)\Psi \left( {\bf y^\prime} \right)   \right\rangle
= \overline \psi \left( {\bf y} \right)\exp \left(
{-ie\int\limits_{\bf y}^{\bf y^\prime} {dz^i A_i \left( z \right)}
} \right)\psi \left( {\bf y^\prime} \right)\left| 0 \right\rangle,
\label{est}
\end{equation}
where $\left| 0 \right\rangle$ is the physical vacuum state and
the integral is to be over the linear spacelike path starting at
$\bf y$ and ending at $\bf y^ \prime$, on a fixed time slice. Note
that the strings between fermions have been introduced to have a
gauge-invariant state $ \left| \Omega  \right\rangle$, in other
terms, this means that the fermions are now dressed by a cloud of
gauge fields. We can write the expectation value of $H$ as
\begin{equation}
\left\langle H \right\rangle _\Omega   = \left\langle \Omega
\right|\int {d^2 } x\left( {\frac{1}{2}E_i^2  + \frac{1}{2}B^2   +
J^i A_i } \right)\left| \Omega  \right\rangle. \label{pot1}
\end{equation}
The preceding Hamiltonian structure thus leads to the following
result
\begin{equation}
\left\langle H \right\rangle _\Omega   = \left\langle H
\right\rangle _0  +  \ \frac{{e^2 }}{2}\int {d^2 x\left(
{\int\limits_{\bf y }^{\bf y^\prime} {dz_i \delta ^{(2)} \left( {x
- z} \right)} } \right)} ^2 , \label{pot2}
\end{equation}
where $ \left\langle H \right\rangle _0  = \left\langle 0
\right|H\left| 0 \right\rangle$. Following our earlier procedure
\cite{Gaete2}, we see that the second term on the right-hand side
of Eq. (\ref{pot2}) is clearly dependent on the distance, and the
potential for two opposite located at $\bf y$ and $\bf y \prime$
takes the form
\begin{equation}
V = \frac{{e^2 }}{{\pi }}\ln \left(\eta |\bf y - \bf y^ \prime|\right) .
\label{pot3}
\end{equation}
This result displays the confining character of the Maxwell
potential for the composites. We see that the result (\ref{pot3}) agrees with
the behavior of the Maxwell-Chern-Simons theory, Eq.(\ref{potz}), in the limit of short separation.

In summary we have shown that while both systems displays
fractional statistics, the composite system proposed here
interacts with purely maxwellian photons and displays a potential
with the confining nature (the potential grows to infinity when
the mutual separation grows). Differently, it is known that use of
the Chern-Simons term turns the electric and magnetic fields
massive leading to a screening potential between static charges.
The analysis above has clearly revealed that, although both
theories lead to fractional statistics by the same mechanism of
attaching a magnetic flux to the charges, their physical contents
are quite different. The observation in the present work that the
composite leads to fractional statistics and confinement is new.

\section{ACKNOWLEDGMENTS}

One of us (CW) would like to thank D. Altbir, J. C. Retamal, J. Gamboa and the Physics Department at USACH for the
invitation and the nice atmosphere that is essential for the
scientific work, he also acknowledges financial suppport from CAPES and CNPq, brazillian research agencies.


\begin{thebibliography}{}

\bibitem{Wilczek} F. Wilczek, Phys. Rev. Lett. {\bf 48},
1144 (1982); {\bf 49}, 957 (1982).

\bibitem{Elcio} E. Abdalla and R. Banerjee, Phys. Rev. Lett. {\bf 80},
238 (1998).

\bibitem{Deser} S. Deser, R. Jackiw, and S. Templeton, Ann. Phys.
(N.Y.) {\bf 140}, 372 (1982).

\bibitem{Dunne} G. Dunne, hep-th/9902115.

\bibitem{Khare} A. Khare, {\it Fractional Statistics and Quantum
Theory } (World Scientific, Singapore, 1998).


\bibitem{Shisuya} K. Shisuya and H. Tamura, Phys. Lett. {\bf B252},
412 (1990).

\bibitem{Itzhaki} N. Itzhaki, Phys. Rev. {\bf D67},
065008 (2003).

\bibitem{Gaete3} P. Gaete, Mod. Phys. Lett. {\bf A19}, in press; hep-th/0305025.

\bibitem{Gaete} P. Gaete and I. Schmidt, Phys. Rev. {\bf D61},
125002 (2000); Phys. Rev. {\bf D64}, 027702 (2001); Phys. Lett.
{\bf B515}, 382 (2001).


\bibitem{Deser-GA} S. Deser, Phys. Rev. Lett. {\bf 64},
611 (1990).

\bibitem{Colatto} For recent investigations on this kind of composite, however in the Maxwell-Chern-Simons model, see for instance; J. Stern, Phys.Lett.B265, 119 (1991) and L.P. Colatto, J.A. Helayel-Neto, M. Hott, Winder A. Moura-Melo; Phys.Lett.A314, 184 (2003) and references therein. 

\bibitem{Dirac1} P. A. M. Dirac, {\it Lectures on Quantum Mechanics}, Belfer Graduate School Monographs,
No. 3, Yeshiva University, 1964.

\bibitem{Dirac} P. A. M. Dirac, {\it The Principles of Quantum
Mechanics} (Oxford University Press, Oxford, 1958); Can. J. Phys.
{\bf 33}, 650 (1955).

\bibitem{Gaete2} P. Gaete, Phys. Rev. {\bf D59},
127702 (1999).


\end{thebibliography}
\end{document}